\documentclass[lettersize,journal]{IEEEtran}
\usepackage{amsmath,amsfonts}
\usepackage{algorithmic}

\usepackage{array}
\usepackage[caption=false,font=normalsize,labelfont=sf,textfont=sf]{subfig}
\usepackage{textcomp}
\usepackage{stfloats}
\usepackage{url}
\usepackage{verbatim}
\usepackage{graphicx}
\usepackage{cite}

\usepackage[ruled,vlined,linesnumberedhidden]{algorithm2e}
\begin{document}
	
	\title{A Novel SCL Bit-Flipping Decoding Of Polarization-Adjusted Convolutional (PAC) Codes}
	\author{
		\IEEEauthorblockN{Wei Zhang}\\
		\IEEEauthorblockA{Nanjing University of Posts and Telecommunications, Nanjing 210003, China}\\
		\IEEEauthorblockA{1021010419@njupt.edu.cn}
	}
	
	\maketitle
	\begin{abstract}
		Polar codes have attracted the attention of numerous researchers in the past decade due to their excellent performance. However, their performance at short block lengths under standard successive cancellation decoding is far from desirable. An effective method to improve the performance at short lengths is CRC precoding followed by successive-cancellation list decoding. Later, Arikan presented polarization-adjusted convolutional (PAC) codes, which further improve the performance of polar codes. In fact, bit-flipping is another post-processing method that can improve decoding performance. In this paper, we propose a novel SCL Bit-Flipping of PAC Codes. We show that better performance can be achieved using list decoding when the list size is the same for PAC codes (N=128, K=64). The decoding performance of our newly proposed PAC-SCLF with a list size of 32 is 0.3 dB better than that of the traditional PAC-SCL with a list size of 32. We set the maximum number of bit flips to 5. The performance of the list size (L=32) for PAC-SCLF is almost the same as the performance of the list size (L=128) for PAC-SCL.
	\end{abstract}
	
	\section{Introduction}
	In 2009, Arikan proposed a kind of channel codes called polar codes which have attracted lots of attention from academic community and industry due to their excellent performance \cite{Ref_1}. In binary-input memory-less symmetric channels, polar codes can achieve the Shannon capacity under successive cancellation (SC) decoder with infinite code length. However, for finite code length, the frame error rate (FER) performance of polar codes under SC decoder is suboptimal. To remedy this deficiency, successive cancellation list (SCL) decoder \cite{Ref_2} and cyclic redundancy check (CRC) aided SCL decoder named CA-SCL decoder \cite{Ref_3} were proposed. Along with the list method, various improvements have remained the statement of article in \cite{Ref_4} - \cite{Ref_6}. 
	
	Besides the article previously covered polar code decoding algorithms, SC Flipping algorithm is firstly proposed in \cite{Ref_7}, where an error-prone information bit will be flipped during next decoding attempt as long as present decoded result cannot pass CRC check. Int this way the FER performance is dramatically improved compared with SC decoding \cite{Ref_1}. In order to reduce the search scope of flipping bits, CRC mechanism \cite{Ref_8}, \cite{Ref_9} was referred in, where polar codeword is divided into segments and each of them are concatenated with a couple of CRC bits. In addition, a family of re-decoding schemes, called SCL-Flip decoding \cite{Ref_10}, \cite{Ref_11}, was proposed for CA-SCL decoding schemes. In each re-decoding attempt in SCL-Flip decoding method, the decision for the CA-SCL decoding on the path competition for a specially selected information bit is alternated, and then standard SCL decoding is conducted for the remaining bits.
	
	At the ISIT in 2019, Arikan proposed a significant breakthrough in polar coding, which boosts the performance of polar codes at short lengths even further. Specifically, a new polar coding scheme \cite{Ref_12}, which he calls polarization-adjusted convolutional codes. At low SNR, the FER performance of PAC-SCL decoding is very close to the BIAWGN dispersion bound approximation. In \cite{Ref_13}, the results show that PAC codes are superior to polar codes and Reed-Muller codes, and the goal of rate-profiling may be to optimize the weight distribution at low weights. \cite{Ref_14} -\cite{Ref_17}, further improve the performance of PAC.
	
	In fact, the SCL decoding procedure of PAC is almost like with compare traditional SCL decoding procedure of polar codes, which is just one more convolutional module before encoding and after decoding. Therefore, bit-flipping is effective for PAC-SCL decoding. 
	
	In this letter, we firstly put forward bit flipping applied in PAC-SCL decoding, and the performance was greatly improved compared with the traditional PAC-SCL decoding. When the RM code was used to construct the polar codes, N=128, K=64, and L=32, the FER performance was increased by 0.3 dB.
	 
	\begin{figure}[ht]
		\centering
		\includegraphics[width=90mm]{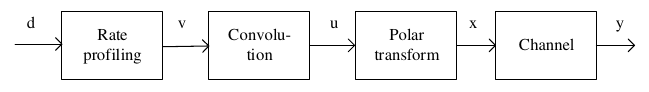}
		\caption{PAC encoding scheme}
	\end{figure}

	\section{Preliminaries}
	Polarization-adjusted convolution codes are convolutional precoding which is performed before the polar codes. The pre-transformation is performed by a rate-1 convolutional encoding as shown Fig. 1. In this section, we first review polar codes and list decoding, SCF and SCLF, then we will focus on PAC codes.
	\subsection{Polar Codes and List Decoding}
	Consider a $(N, K, \mathcal{A})$  polar code with block-length $N=2^n$, information bit length $K$ and information bit index set of $\mathcal{A}$.  Let  $u_1^N=(u_1,u_2,...,u_N)$  denote the input vector to be encoded, where $u_i$  is an information bit whenever  $i\in \mathcal{A}$ and a frozen bit for  $i\in \mathcal{A}^c$.  For polar encoding, $x_1^N=u_1^N\times G_N$ is employed with $G_N=F^{\otimes n}$, where $F^{\otimes n}$ denotes the $n$-th Kronecker power of $F=\left[ {\begin{array}{*{20}{c}}
			1&1\\
			1&0
	\end{array}} \right]$.
	
	For SC decoding, it  successively evaluates the log-likelihood ratio of each bit $u_i$ based on the received vector $y_1^N$ and its $i$ preceding decision bits $ \hat{u}_{1}^{i-1}$
	\begin{eqnarray}
		L^i = \log \frac{P\left(y_1^{N},\hat u_1^{i - 1}|{u_i} = 0\right)}{P\left(y_1^{N},\hat u_1^{i - 1}|{u_i} = 1\right)}.
	\end{eqnarray}
	Then,  $\hat{u}_i$ is decided as 0 if $L^i\geq 0$ and as 1 if $L^i\leq 0$. 
	
	Instead of just keeping  a single path, SCL preserves $L > 1$ paths during the decoding process, which could significantly improve the decoding performance. Let  $L_l^i$ denote the log-likelihood ratio of the bit $u_i$ along the $l$-th path. When the number of paths is greater than the list size as the SCL decoder proceeds from  level $i$ to $i+1$, it retains $L$ best paths according to the updated  path metric
	\begin{flalign}
		PM_l^{i} =\left\{
		\begin{aligned}
			&PM_l^{i - 1},\quad \quad \quad \: {\text{if} \: {\hat u}_{i,l}} = \frac{1}{2}[1 - \text{sign}(L_l^i)]\\ 
			&PM_l^{i - 1} + |L_l^i|,\: \text{otherwise}.
			\\
		\end{aligned}
		\right.
	\end{flalign}
	where  $\text{sign}(x )=1$ if $x>0$ and -1 otherwise. 
	
	After all nodes  in  $\mathcal{A}$ are visited, the path with the smallest path metric is selected as the survival path.  For CA-SCL decoding,  the output $L$-paths are checked by CRC. Once a path passes the CRC check, it is claimed as the decoded output. Otherwise, a decoding failure is claimed.
	\subsection{SC-Flip and SCL-Flip Decoding}
	SC-Flip decoding algorithm attempts to flip a decision to get the correct decoding whenever the conventional SC decoding fails \cite{Ref_18}. It was observed that error propagation occurs frequently in SC decoding, where any single erroneous decision may result in a burst of errors. Hence, it is crucial to find the first error position in SC-Flip decoding.
	
	For SCL decoding,  the bit-flipping can be again employed whenever a decoding failure is claimed.  The so-called SCL-Flip decoding was recently proposed in \cite{Ref_10}, \cite{Ref_11}.  In \cite{Ref_10}, a critical set is detected, which is deemed to be error-prone during SCL decoding. Therefore,  each bit in this critical set is flipped in the re-decoding attempts. In \cite{Ref_19}, the bit position for flipping is determined by a newly-introduced confidence metric for the survival paths and simulations shown  significant performance improvement over the scheme in \cite{Ref_10}.
	\subsection{Brief overview of PAC codes }
	PAC codes are concatenated codes in which a convolutional transform is employed before polar encoding. Polar codes the block length N of PAC code is also $2^n$. It's shown Fig. 1 that the process of PAC encoding consists of three parts: rating-profiling, convolution precoding and polar transform. 
	
	The information bits $\textbf{d}=(d_0,d_1,...,d_{K-1})$ are first mapped to a vector $\textbf{v}=(v_0,v_1,...,v_{N-1})$ using a rate-profile (RM-construction \cite{Ref_20} ). The rate-profile is formed based on the index $\mathcal{A}$ such as $u_{\mathcal{A}}=d$ and $u_{\mathcal{A}^c}=0$. Meanwhile, $u_{\mathcal{A}^c}=0$ represents frozen bits and the other represents information bits. After rate-profiling, the vector $\mathbf{v}$ is transformed using a convolutional generator polynomial $\mathbf{g}=[g_0,g_1,...,g_m]$ to $u_i=\sum_{j=0}^{m}g_jv_{i-j}$ (more description sees subroutine conv in Algorithm 1), where $g_i\in \{0,1\}$. In this letter, we take advantage of convolutional generator polynomial which is $\mathbf{g}=\{1,0,1,1,0,1\}$. In summary, the polar trans-formation is performed by $\mathbf{x}=\mathbf{u} \cdot \mathbf{G}$, $\mathbf{G}$ represents the generator matrix of polar codes.
	
	\begin{algorithm}[h]
		\caption{PAC encoding}
		\KwIn{profiling information $\mathbf{v}$,  generator polynomial $\mathbf{g}$}
		\KwOut{the codeword $\mathbf{x}$}
		\SetKwFunction{Fsubroutine }{subroutine $conv(\mathbf{v}, \mathbf{g})$}
		\SetKwFunction{Fsubroutine }{subroutine $subconv(v, curState,  \mathbf{g})$}
		$\mathbf{u}$ $\leftarrow$ $conv$($\mathbf{v}$, $\mathbf{g}$) \quad // conv means convolutional encoder \\
		$\mathbf{x}$ $\leftarrow$ $enc$($\mathbf{u}$) \quad \quad // enc means polar encoder\\
		\Return $\mathbf{x}$;\\
		
		\SetKwFunction{Fsubroutine}{subroutine $conv(\mathbf{v}, \mathbf{g})$}
		\SetKwProg{Fn}{}{:}{\KwRet}
		\Fn{\Fsubroutine}{
			$curState$$[g_0,g_1,...,g_{m-1}]$ $\leftarrow$ $[0,0,...,0]$\\
			\For{i $\leftarrow$ 0 to $|\mathbf{v}|$-$1$}
			{$(u_i, curState)$ $\leftarrow$ $subconv$($v_i, curState, \mathbf{g}$)
			}
			\Return $\mathbf{u}$;
		}
	
		\SetKwFunction{Fsubroutine}{subroutine $subconv(v, curState,  \mathbf{g})$}
		\SetKwProg{Fn}{}{:}{\KwRet}
		\Fn{\Fsubroutine}{
			$temp$ $\leftarrow$ v \\
			\For {$j$ $\leftarrow$ 0 to $|g|$-$1$}{
				$v$ $\leftarrow$ $v$ $\oplus$ $curState$ $[j]$
			}
			$nextState$ $\leftarrow$ $curState$$[temp,g_0,g_1,...,g_{m-2}]$\\
			\Return $(v,nextState)$;
		}
	\end{algorithm}

	\section{Bit-flipping List Decoding Of PAC Codes}
	In this section, we review the SCL decoding of PAC. The SCL decoding scheme of PAC codes is shown in Fig. 2.  Then, we will elaborate on our contribution that constructs a bit-flipping set during SCL decoding of PAC. This operation greatly improves the decoding performance. So it is necessary to introduce the application of bit-flipping in SCL decoding of PAC in detail.
	
	\begin{figure}[ht]
		\centering
		\includegraphics[width=90mm]{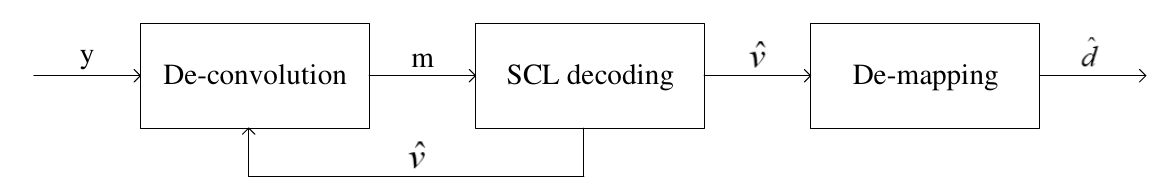}
		\caption{SCL decoding of PAC codes scheme}
	\end{figure}
	\subsection{SCL decoding of PAC codes}
	SCL decoding of PAC can be regarded as the convolutional decoding embedded in the traditional SCL decoding of polar codes. More details will be introduced in Algorithm 2. Our implementation is similar to \cite{Ref_21}. We discover the list decoding for PAC codes which trades a fixed time complexity for a large memory requirement (to store a list
	of paths) and is easier to implement. In the context of PAC codes, we reproduced the results of article \cite{Ref_13}. 
	
	 Algorithm 2 shows the list decoding approach. Before arriving at the first information bits, it is only a single path in the list. The decoder knows the value of frozen bits, thus $v_i=0$, then according to the current memory state cur and the generator polynomial $\mathbf{g}=x^6+x^4+x^3+x+1$. The function $subconv$ is identical with the one in Algorithm 1. Take advantage of the function f or function g to calculate the LLR values. The corresponding path metric is calculated using subroutine $calcPM$. When the value $u_i$ is known,  We can calculate the value of the partial sum using $updateSums$.
	
	\begin{algorithm}[h]
		\caption{List decoding of PAC codes}
		\KwIn{channel LLRs $llr_0^{N-1}$, $\mathcal{A}$, L, $\mathbf{g}$}
		\KwOut{message bits $\mathbf{\hat{d}}$}
		\SetKwFunction{Fsubroutine }{subroutine $conv(\mathbf{v}, \mathbf{g})$}
		\SetKwFunction{Fsubroutine }{subroutine $subconv(v, curState,  \mathbf{g})$}
		
		$L \leftarrow {1}$ \quad \quad // initial only a single path\\
		$[llr,p] \leftarrow [llr_0^{N-1},0]$ \quad // $p$ means partial-sums\\
		\For{$i \leftarrow 0 $ to $N-1$}{
			// frozen bits\\
			\If{$i \in \mathcal{A}^c$ }{
				\For{$l \leftarrow 1$ to $|L|$}{
				$llr_0^i[l] \leftarrow updateLLRs(l,i,llr[l],p[l])$	\\
				$\hat{v}_i[l] \leftarrow 0$\\
				$[\hat{u}_i[l],cur[l]] \leftarrow subconv(\hat{v}_i[l],cur[l],g)$\\
				$PM_l^{i} \leftarrow calPM(PM_l^{i-1},llr_0^i[l],\hat{u}_i[l])$\\
				$p[l] \leftarrow updateSums(\hat{u}_i[l],p[l])$
			}
			}
			\Else{
				\For{$l \leftarrow 1$ to $|L|$}{
				// duplicate-path	\\
				$L \leftarrow dulPath(L,l,i,g)$	
				}
			
			\If{$L >$  $\rm{L}$}{
				// delete-path \\
				$L \leftarrow delPath(L)$
			}
			}
		}
		// $demapping$ means transforming receiving bits into a\\// message bits
		\\
		$\hat{\textbf{d}} \leftarrow demapping(\hat{v}_1^N[0])$\\
		\Return $\hat{\textbf{d}}$;\\
		
		\SetKwFunction{Fsubroutine}{subroutine $dulPath(L,l,i,g)$	}
		\SetKwProg{Fn}{}{:}{\KwRet}
		\Fn{\Fsubroutine}{
			$llr_0^i[l] \leftarrow updateLLRs(l,i,llr[l],p[l])$\\
			$\hat{v}_i[l] \leftarrow 0$, 
			$\hat{v}_i[l'] \leftarrow 1$\\
			$[\hat{u}_i[l],cur[l]] \leftarrow subconv(\hat{v}_i[l],cur[l],g)$\\
			$[\hat{u}_i[l'],cur[l']] \leftarrow subconv(\hat{v}_i[l'],cur[l'],g)$\\
			$PM_l^{i} \leftarrow calPM(PM_l^{i-1},llr_0^i[l],\hat{u}_i[l])$\\
			$PM_{l'}^{i} \leftarrow calPM(PM_{l'}^{i-1},llr_0^i[l'],\hat{u}_i[l'])$\\
			$p[l] \leftarrow updateSums(\hat{u}_i[l],p[l])$\\
			$p[l'] \leftarrow updateSums(\hat{u}_i[l'],p[l'])$\\
			$L \leftarrow l \cup l' $\\
			\Return $\textbf{L}$
		}
		
		\SetKwFunction{Fsubroutine}{subroutine $calPM(PM, llr,  \mathbf{g})$}
		\SetKwProg{Fn}{}{:}{\KwRet}
		\Fn{\Fsubroutine}{
			\If{$\hat{u}$ = $\frac{1}{2}(1-sign(llr))$}{
				
				$PM=PM$;
			}
			\Else{
				$PM=PM+|llr|$;
			}
		\Return $\textbf{PM}$
		}
	
		\SetKwFunction{Fsubroutine}{subroutine $delPath(L)$}
		\SetKwProg{Fn}{}{:}{\KwRet}
		\Fn{\Fsubroutine}{
		according to the calculated PM value\\ 
		L paths with smaller path metrics are retained\\
		 L paths with larger path metrics are deleted
		}
	\end{algorithm}
	\noindent On the other hand, if the index of the current bit is in the set $\mathcal{A}$, there exists two options for value of $v_i$ 0 and 1. For each option of 0 and 1, the process for $\mathcal{A}$ including convolutional encoding, calculating path metric the encoded values $u_i=0$ and $1$ are fed back into SCL process. the two encoded values ui = 0
	and 1 are fed back into SC process. The subroutines updateLLRs, updateSums, delPath already are introduced the Algorithm 2. Note that the vector $llr$ and $p$ are the LLRs and parity sums.
	
	It is worth noting that the SCL decoding of PAC is almost the same as the traditional polarization code SCL decoding process, except for additional a process of convolutional re-encoding at each decoding step which needs the next memory state is stated for the next path. To reduce the computational complexity and performance of list decoding, the methods proposed in the literature such as in \cite{Ref_22}, \cite{Ref_23} can be applied to PAC list decoding as well.
	
	In the following section, we will describe our work to improve the SCL decoding algorithm of PAC by applying the bit flipping technique to the decoding algorithm.
	\subsection{Constrcut bit-flipping set of PAC codes}
	In \cite{Ref_19}, it was shown that the decoding failures in the CA-SCL decoding are mainly caused by the elimination of the correct path from $L$ maintaining paths. In general, bit-flipping, as a post-processing technique, could be repeatedly implemented if the previous attempt fails. It is worth considering whether applying bit flipping to SCL decoding of PAC would be an improvement, since the SCL decoding principle of PAC is similar to the traditional SCL decoding principle of polar codes. Therefore, we attempt to use this idea and achieve performance boost.
	
	At the beginning of SCL bit-flipping decoding of PAC, the first thing we need to do is construct the bit-flipping set. According to \cite{Ref_19}, it was shown that the confidence in the decision for the path competition on $u_i,  i \in \mathcal{A}\setminus {\mathcal{A}_0}$, can be determined
	from the ratio between the total probability of the $L$ survival paths to the total probability of the $L$ removed paths, namely, 
	\begin{eqnarray}
		\label{eq:conf}
		{E_i}(\alpha ) = \log \frac{{\sum\nolimits_{l = 1}^L {{e^{ - PM_l^{i}}}} }}{{{{\left( {\sum\nolimits_{l = 1}^L {{e^{ - PM_{l + L}^{i}}}} } \right)}^\alpha }}}
	\end{eqnarray}

	in the case of $\alpha\ge 1$. Note that $\mathcal{A}_0$ denote the set consisting of the least $\log_2 L$ information indices. However, once the first decoding failure at level $i_1 \in \mathcal{A}$ occurs during the SCL decoding of PAC, it is likely to produce  error propagation in the subsequent decoding process for $i$ $>$ $i_1$. In this case, there will be a biased estimate in the confidence of the decision which is less than its correct value for a large index $i$. To compensate for the biased estimate due to the error propagation, $\alpha\ge 1$ is introduced in (\ref{eq:conf}). Essentially, an information index $i$ that has a low $E_i(\alpha)$ should have a high priority for re-decoding. Therefore, a bit-flipping index set is constructed in \cite{Ref_19} by locating the bit indexes with smallest values of $E_i(\alpha)$. 

	This may simply adopt the following rule, namely
	\begin{eqnarray}
		\label{eq:min}
		\hat{i}_1 = \min_{i} E_i(\alpha),   i\in \mathcal{A}\setminus \mathcal{A}_0. 
	\end{eqnarray}	

	However, whenever  a decoding failure occurs,  there may exist multiple bit errors in the final decoded output $\hat{u}_1^N$.  This means that  $Q> 1$ error positions $\{i_1,\cdots,i_Q\}\subset \mathcal{A}$ may appear in $\hat{u}_1^N$.  All of these error positions may have low confidence in $E_{i_q}(\alpha), q\in [1,Q]$. This makes the location of the first error position using (\ref{eq:min}) unreliable.	
	
	Then, defining a bit-flipping set as follows:
	\begin{eqnarray}
		\begin{aligned}
			\mathcal{E}^{\alpha}=\{E_i(\alpha)\,|\, i\in \mathcal{A}\setminus \mathcal{A}_0\}, \\
		\end{aligned}
	\end{eqnarray}	

	By sorting the above set independently, the index of elements in sorted ascending order  can be obtained as
	\begin{eqnarray}
		\begin{aligned}
			\mathcal{I}=sort(\mathcal{E}^{\alpha})\\
		\end{aligned}
	\end{eqnarray}	

	where $sort(\cdot)$ returns the indices of $ \mathcal{E}^{\alpha}$ with the ascending order of  $E_{i_1}^{\alpha}\leq E_{i_2}^{\alpha}\leq...\leq E_{i_Q}^{\alpha}$.
	
	\begin{algorithm}[h]
		\caption{  GenFlip bit-flipping set $\mathcal{F}$ }
		\KwIn{$\{E_i{(\alpha)}, i\in \mathcal{A}\setminus \mathcal{A}_0\}$ }

		\KwOut{$\mathcal{F}$}

		$\mathcal{E}^{\alpha}\leftarrow E_i{(\alpha)}$ \\
		$\mathcal{I} \leftarrow sort(\mathcal{E}^{\alpha})$ \\
		\For{$m=1$ to $Q$}{
			$\mathcal{F}\leftarrow \mathcal{I}(1:m)$\\
		} 
		return $\mathcal{F}$\\
		\textbf{end procedure}
	\end{algorithm}
	
	\subsection{SCL bit-flipping decoding of PAC codes}
	When the first SCL decoding of PAC codes is performed, the return value is slightly different from the traditional SCL decoding of PAC codes. There exists additional $E_i(\alpha)$ according to formula ($3$). If the receiving bits $u_1^N[1] \neq \textbf{v}$, we make a judgment that this process SCL decoding of PAC codes fails. Then, the strategy we take is to construct a bit-flipping set according to $GenFlip$ based on Algorithm 3. The more advanced the position of the elements in the bit-flipping set $\mathcal{F}$, the greater the probability that the first information bit will be wrong, so it is necessary to iterate from the first element in $\mathcal{F}$ until the last element of $\mathcal{F}$ is reached, or the receiving bits is the correct decoding result. In algorithm 4, $\mathcal{F}[m]$ means that when an error occurs in the received bit, the current decoding process is restarted, which is equivalent to the position of the information bit that needs to be flipped. In subroutine $\textbf{\rm{PAC}}$-$\textbf{\rm{SCLF}}$. Only the information bit position that needs to be flipped is shifted \cite{Ref_24}, and other information bits and frozen bits are used for normal SCL decoding of PAC codes.
	\begin{algorithm}[h]
		\caption{Bit-flipping list decoding of PAC codes}
		\KwIn{channel LLRs $llr_0^{N-1}$, $\mathcal{A}$, L, $\mathbf{g}$}
		\KwOut{message bits $\mathbf{\hat{d}}$}
		\SetKwFunction{Fsubroutine }{subroutine $conv(\mathbf{v}, \mathbf{g})$}
		\SetKwFunction{Fsubroutine }{subroutine $subconv(v, curState,  \mathbf{g})$}
		
		$(u_1^N[1], E_i{(\alpha)} )\leftarrow$ PAC-SCL$($$llr_0^{N-1}, \mathcal{A},$ L, g$)$\\
			
		\If{$u_1^N[1]=\textbf{v}$}{
			${\hat{\textbf{d}}} \leftarrow demapping(u_1^N[1])$
		}
		\Else{
			$\mathcal{F} \leftarrow  GenFlip({E_i(\alpha)})$\\
			\For{$m=1$ to $Q$}{
				$u_1^N[1] \leftarrow$  PAC-SCLF$($$llr_0^{N-1}, \mathcal{A},$ L, g, $\mathcal{F}[m])$\\
				\If{$u_1^N[1]=\textbf{v}$}{
					${\hat{\textbf{d}}} \leftarrow demapping(u_1^N[1])$\\
					\textbf{break}
				}
				\Else{
					\If{m=Q}{
					${\hat{\textbf{d}}} \leftarrow demapping(u_1^N[1])$\\	
					}
				}
			}
		}
	\Return ${\hat{\textbf{d}}}$
	
	\SetKwFunction{Fsubroutine}{subroutine PAC-SCLF$($$llr,$$\mathcal{A}$$,$L$,$g$,$$index)$}
	\SetKwProg{Fn}{}{:}{\KwRet}
	\Fn{\Fsubroutine}{
		\For{$n\leftarrow 1\, to\, N$}{
			Perform one standard PAC-SCL Decoding:\\
			Path pruning when $n\in \mathcal{A} $ AND $L<2^n$\\
			\If{$n\neq index$}{
				$List\leftarrow \{1,...,L\}$\\
			}
			\Else{
				$List\leftarrow \{L$+$1,...,2L\}$\\					
			}
		}
		\Return $\hat{u}_1^N[1]$
	}

	\end{algorithm}
	\section{Simulations}
	
	In this section, simulations are performed to show the effectiveness of the proposed SCL bit-flipping decoding of PAC codes (PAC-SCLF), which compares with traditional SCL decoding of PAC codes (PAC-SCL) \cite{Ref_13}. We use the code length N which is 128 and the information bits length K which is 64, the generator polynomial of the convolution module is $g=x^6+x^4+x^3+x^1+1$.The polar codes are constructed by RM code, all the code bits are BPSK $(1\to-1, 0\to1)$ modulated and transmitted over an AWGN channel ($\sigma=\frac{1}{\sqrt{2R}}10^{-\frac{snr}{20}}$).

	\begin{figure}[ht]
		\centering
		\includegraphics[scale=0.6]{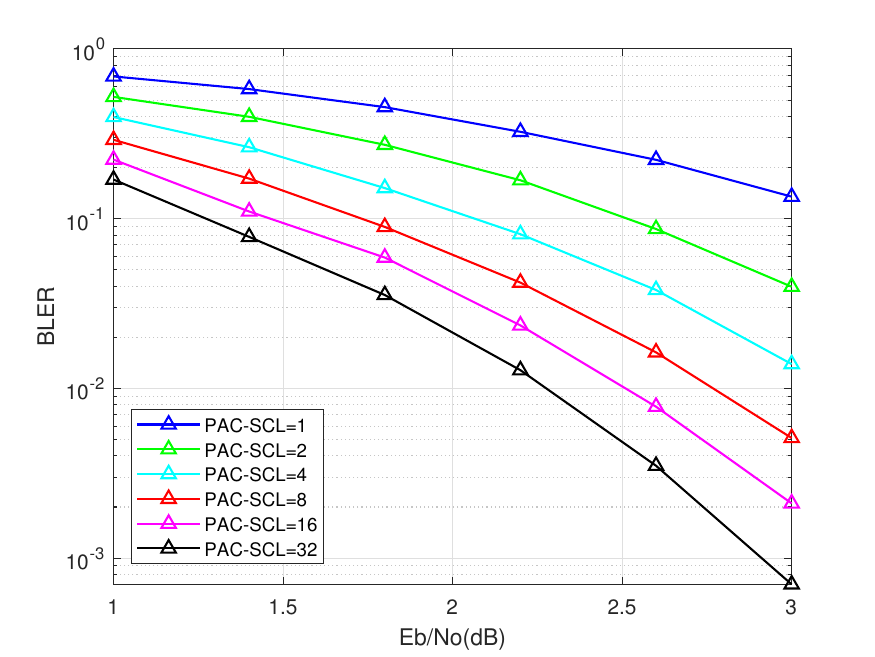}
		\caption{BLER performance SCL decoding of PAC codes  with different list size}
		\label{fig:ListSize}
	\end{figure}

	Fig. 3 shows the performance of the traditional SCL decoding of PAC codes (128,64). The construction used in polar codes is known as Reed-Muller (RM) code construction. As the list grows, the performance of decoding improves.

	\begin{figure}[ht]
		\centering
		\includegraphics[scale=0.6]{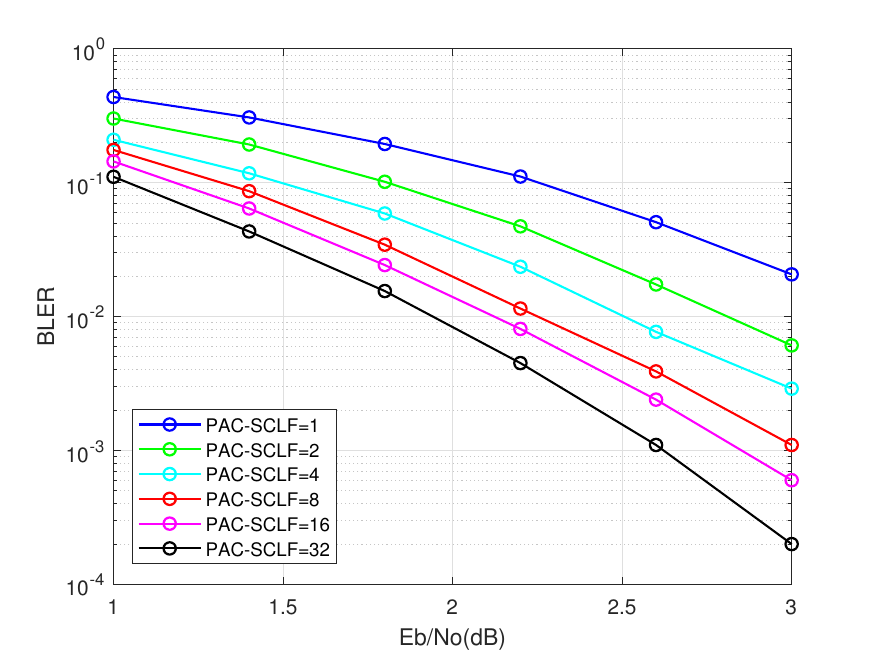}
		\caption{BLER performance SCLF decoding of PAC codes  with different and list size}
		\label{fig:ListSize}
	\end{figure}

	Fig. 4 shows a significant improvement in performance of our proposed SCL bit-flipping decoding based on PAC codes (128,64). We set the maximum number of bit-flipping as T$=$5. Based on Algorithm 4, if the maximum bit-flipping count T is not reached during this process and the decoding result is already correct, it exits directly. Otherwise, the bit-flipping count needs to reach the maximum T in order to consider it as finished.
	
	\begin{figure}[ht]
		\centering
		\includegraphics[scale=0.6]{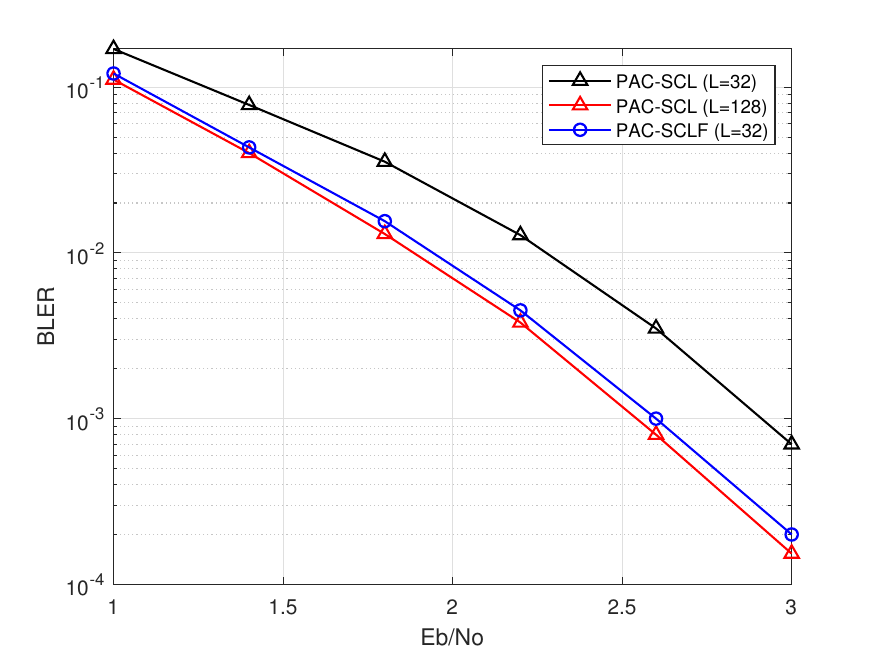}
		\caption{BLER performance comparison for SCL decoding of PAC codes and SCLF decoding of PAC codes}
		\label{fig:ListSize}
	\end{figure}

	Fig. 5 represents the performance comparison of our proposed decoding scheme PAC-SCLF decoding and the traditional PAC-SCL decoding algorithm for PAC code (128,64) with a list size of 32. Our proposed scheme shows an improvement of approximately 0.3dB compared to the traditional PAC-SCL decoding algorithm. Moreover, the performance is similar to using the traditional PAC-SCL decoding algorithm with a list size of 128.

	\section{Conclusion}
	This article introduces for the first time the application of bit-flipping strategy in the PAC-SCL decoding process. It is indeed an interesting approach. The relatively low number of flips we set leads to reduced latency in the decoding process. Simulations show that the proposed PAC-SCLF could achieve obvious improvement over the standard PAC-SCL decoding.

\end{document}